# THE FORMATION OF STELLAR BLACK HOLES

I.F. Mirabel[1,2*]

1. Institute of Astronomy and Space Physics. CONICET-Universidad de Buenos Aires. Ciudad Universitaria, 1428 Buenos Aires, Argentina
2. Laboratoire AIM-Paris-Saclay, CEA/DSM/Irfu−CNRS, CEA-Saclay, pt courrier 131, 91191 Gif-sur-Yvette, France
* Correspondence to Félix Mirabel: [felix.mirabel@cea.fr](felix.mirabel@cea.fr)

**Abstract:** It is believed that stellar black holes (BHs) can be formed in two different ways: Either a massive star collapses directly into a BH without a supernova (SN) explosion, or an explosion occurs in a proto-neutron star, but the energy is too low to completely unbind the stellar envelope, and a large fraction of it falls back onto the short-lived neutron star (NS), leading to the delayed formation of a BH. Theoretical models set progenitor masses for BH formation by implosion, namely, by complete or almost complete collapse, but observational evidences have been elusive. Here are reviewed the observational insights on BHs formed by implosion without large natal kicks from: (1) the kinematics in three dimensions of space of five Galactic BH X-ray binaries (BH-XRBs), (2) the diversity of optical and infrared observations of massive stars that collapse in the dark, with no luminous SN explosions, possibly leading to the formation of BHs, and (3) the sources of gravitational waves produced by mergers of stellar BHs so far detected with LIGO. Multiple indications of BH formation without ejection of a significant amount of matter and with no natal kicks obtained from these different areas of observational astrophysics, and the recent observational confirmation of the expected dependence of BH formation on metallicity and redshift, are qualitatively consistent with the high merger rates of binary black holes (BBHs) inferred from the first detections with LIGO.

## 1. Introduction

The formation of stellar BHs is of topical interest for several areas of astrophysics. Stellar BHs are remnants of massive stars, possible seeds for the formation of supermassive BHs, and sources of the most energetic phenomena in the universe, such as the gravitational waves produced by fusion of BHs.

BHs and NSs are the fossils of stars with masses above ~8 $M_\odot$. It is known that some fraction of NSs have large runaway motions, probably due to strong natal kicks (NKs) imparted to the compact object. NKs have also been invoked in models of the core collapse of massive stars that lead to the formation of BHs. Such models predict in addition, that under specific conditions, BHs can also be formed by implosion with no energetic kicks[1,2], depending on mass, binarity, metallicity, angular momentum, and magnetic fields, among others properties of the progenitor star. NKs are of interest in Gravitational Wave Astrophysics since from population synthesis models of isolated binary evolution it is inferred that the merger rate of BBHs increases by a factor of ~20 when BH NKs are decreased from a kick distribution typical of NSs to zero[3].

It is believed that the runaway velocity of a BH-XRB can be due to the following mechanisms. (1) The sudden baryonic mass-loss in the SN explosion of the primary star of a binary[4] (Blaauw kick). In this case the ejected matter will continue to move with the orbital velocity of the progenitor, and to conserve momentum the resulting compact binary will move in the opposite direction[5]. A sudden mass loss would unbind the binary only when more than half the binary's total mass is instantaneously lost, which is not expected[5]. (2) NKs can also be imparted to the compact object, by anisotropic emission of neutrinos[6] and GWs[7] during core-collapse. If formed in a dense stellar cluster, other possible causes for the runaway velocity of a compact BH-XRB could be either one of several possible dynamical interactions in the stellar cluster[8,9,10], or the explosion of a massive star that before its collapse formed a multiple bound system with the runaway compact binary.

## 2. Kinematics of Galactic black hole X-ray binaries

The kinematics of BH-XRBs can provide clues on the formation of BHs. If a compact object is accompanied by a mass-donor star in an X-ray binary, it is possible to determine the distance, radial velocity, and proper motion of the system's barycenter, from which can be derived the velocity in three dimensions of space, and in some cases the path to the site of birth may be tracked.

Among the estimated $3\times10^8$ stellar BHs in the Galaxy[11], about 20 BH-XRBs have been dynamically confirmed, and until present for only five of those BH-XRBs it was possible to determine their velocity in the three dimensions of space. High-precision astrometric observations with Very Long Baseline Interferometry (VLBI) at radio wavelengths provides model-independent distances from geometric parallaxes, from which can be gathered insights on the X-ray binary systems, and on the formation mechanism of BHs[12]. In Table 1 are listed the known parameters of these five black hole binary systems and the estimated peculiar velocities relative to their local environment and/or birth place.

### 2.1. Black holes formed by implosion

The end of massive stellar evolution depends on metallicity[1,13], binarity[14], angular momentum[15], nucleosynthesis and neutrino transport for the explosions[16], among other possible factors. Despite multiple uncertainties, most models predict that massive stars may collapse forming BHs directly, when no proto-neutron star is formed, a transient proto-neutron star is formed but unable to launch a SN shock, and a black hole is formed by fallback of mass after an initial SN shock has been launched. Stellar BHs may be formed with or without explosive ejection of a significant amount of baryonic matter, and with or without natal kicks.

In Figure 1 is shown as an example what sort of explosion and remnants of single stars are left as a function of initial mass and metallicity of the stellar progenitors, assuming stellar winds are the only means of mass loss[13]. For stars between metal-free and about solar metallicity this model predicts that stars between 25 and 40 $M_\odot$ end as BHs by SN mass fallback, or by direct core collapse when the stellar mass is above ~40 $M_\odot$ and metallicity less than ~0.6 solar.

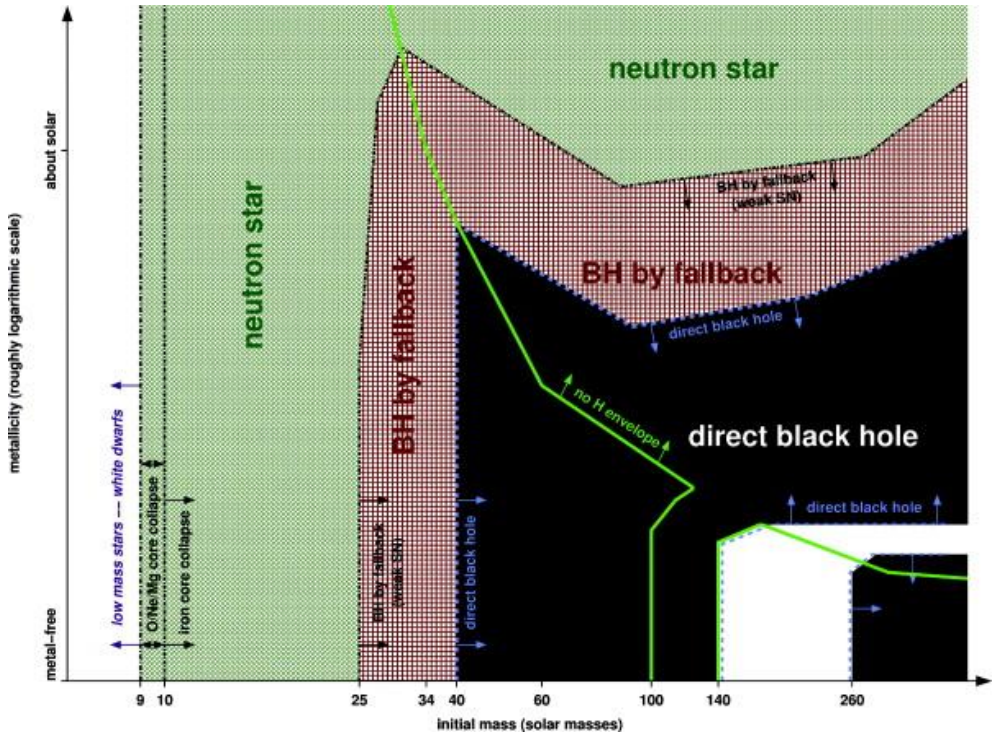

**Figure 1.** Remnants of massive single stars as a function of initial metallicity (*y-axis*; qualitatively) and initial mass (*x-axis*). The thick green line separates the regimes where the stars keep their hydrogen envelope (left and lower right) from those where the hydrogen envelope is lost (upper right and small strip at the bottom between 100 and 140 $M_\odot$. The dashed blue line indicates the border of the regime of direct black hole formation (*black*). This domain is interrupted by a strip of pair-instability supernovae that leave no remnant (*white*). Outside the direct BH regime, at lower mass and higher metallicity, follows the regime of BH formation by fallback (*red cross-hatching and bordered by a black dot-dashed line*). Outside of this, green cross-hatching indicates the formation of neutron stars. The lowest mass neutron stars may be made by O/Ne/Mg core collapse instead of iron core collapse (*vertical dot-dashed lines at the left*). At even lower mass, the cores do not collapse and only white dwarfs are made (*white strip at the very left*). Reproduced from ref. 13.

**Cygnus X-1** is an X-ray binary at a distance of 1.86 ± 0.1 kpc[17] composed of a BH of 14.8 ± 1.0 $M_\odot$ and a O9.7Iab donor star of 19.2 ± 1.9 $M_\odot$ with an orbital period of 5.6 days and eccentricity of 0.018 ± 0.003.

There have been several efforts to model the evolution of Cygnus X-1. Assuming Cygnus X-1 has a runaway velocity of ~50 km s$^{-1}$ it had been estimated that this velocity could be accounted by the ejection of at least 2.6 $M_\odot$ in a successful SN, without need of a NK[5]. In a more recent model[18] it was found that at the time of core collapse the BH received a kick of < 77 km s$^{-1}$ at 95% confidence, but this constraint is not particularly strong.

Later, it was pointed out that Cygnus X-1 appears to be at comparable distance and moving together with the association of massive stars Cygnus OB3, and therefore was proposed[19] that the BH in Cygnus X-1 was formed in situ and did not receive an energetic trigger from a NK or mass loss in a SN. The more recent VLBI measurements at radio wavelengths[17] of the parallax and proper motion of Cygnus X-1, and the new reduction of the Hipparcos data for the association Cygnus OB3[20], show that within the errors, the distance and proper motion, -as well as the radial velocity of the BH X-ray binary barycenter-, are consistent with those of the association of massive stars, which reaffirmed the conjecture that Cygnus OB3 is the parent association of Cygnus X-1. The motions on the plane of the sky of Cygnus X-1 and Cygnus OB3 are shown in Figure 2.

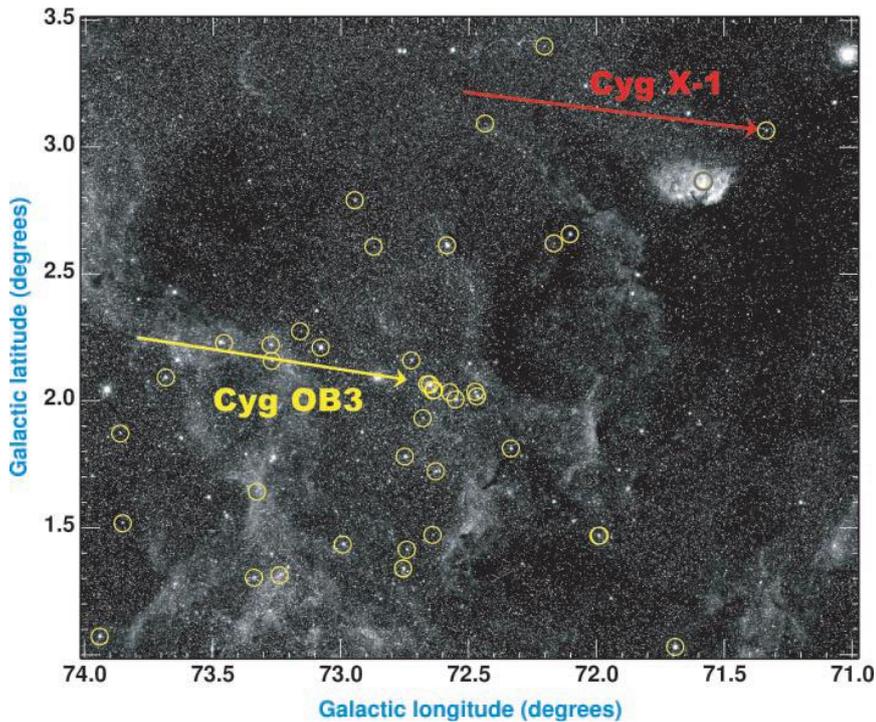

**Figure 2.** Optical image of the sky around the black hole X-ray binary Cygnus X-1 and the association of massive stars Cygnus OB3.The red arrow shows the magnitude and direction of the motion in the plane of the sky of the radio counterpart of Cygnus X-1 for the past 0.5 million years. The yellow arrow shows the magnitude and direction of the average Hipparcos motion of the massive stars of Cygnus OB3 (circled in yellow) for the past 0.5 million years. Despite the different observational techniques used to determine the parallax distances and proper motions, Cygnus X-1 moves in the sky as Cygnus OB3. At a distance of 1.9 kpc, the space velocity of Cygnus X-1 relative to that of Cygnus OB3 is < 9 ± 2 km/s. (Adapted from ref. 19)

The upper limit of the velocity in three dimensions of Cygnus X-1 relative to the mean velocity of Cygnus OB3 is 9 ± 2 km s$^{-1}$, which is typical of the random velocities of stars in expanding stellar associations[4]. The peculiar motion of Cygnus X-1 relative to Cygnus OB3 implies that the X-ray binary would have reached its projected distance of ~56 pc from the center of Cyg OB3 in (6.5 ± 2) x 10$^6$ years[19].

A lower limit for the initial mass of the BH progenitor can be estimated by assuming that all massive stars of the parent stellar association, including the BH progenitor of Cygnus X-1, were formed over a short time span. The highest-mass main-sequence star in Cyg OB3 is of spectral type O7 V and has 40 ± 5 M$_\odot$[21]. Because more massive stars evolve faster, the lower limit for the initial mass of the BH progenitor in Cygnus X-1 is 40 ± 5 M$_\odot$. The upper limit for the initial mass would be equivalent to that of the highest mass stars found in Galactic associations, up to ~100 $M_\odot$. The time since the formation of Cygnus OB3 and the progenitor of Cygnus X-1 as inferred from models of stellar evolution is $(5 \pm 1.5) \times 10^6$ years, and it is consistent with the $(6.5 \pm 2) \times 10^6$ years Cygnus X-1 would have taken to move from the center of Cygnus OB3 to its present position[19].

From the equations for spherical mass ejection in BH formation[5] it is estimated that the maximum baryonic mass that could have been suddenly ejected to accelerate the binary to a velocity of $(9 \pm 2)$ km s$^{-1}$, with no NK, is less than $1 \pm 0.3$ $M_\odot$[19]. Indeed, there are no observational evidences for a SN remnant in the radio continuum, X-rays, and atomic hydrogen surveys of the region where Cygnus X-1 was most likely formed. Before complete collapse the binary progenitor of Cygnus X-1 is likely to have gone through an earlier mass transfer episode, which may not have been conservative, and additional mass may have been lost by stellar winds in a Wolf-Rayet stellar stage. The primary star would have lost by transfer mass episodes and stellar winds a total mass of $> 25 \pm 5$ $M_\odot$ because the initial mass of the progenitor was ~$40 \pm 5$ $M_\odot$, and the estimated BH mass is ~$15 \pm 1.0$ $M_\odot$.

From the kinematics of Cygnus X-1 and its association with Cygnus OB3 it is inferred that the BH was formed by implosion of a star of at least ~40 $M_\odot$. As shown in figure 1, this observational mass lower limit of ~40 $M_\odot$ is the same as that theoretically predicted[22] for the transition from partial fallback to direct collapse of BH progenitors of solar metallicity.

By constraining $V_{pec}$ to <10 km s$^{-1}$, as determined by the observations[19], from model[18] it is found that the helium core had a mass $M_{He}$ in a range of 13.9–16.9 $M_\odot$, at 95% confidence, which is consistent with a remnant BH of ~15 $M_\odot$ resulting from a complete or almost complete collapse of the helium core. For the more recently determined BH mass of ~15 $M_\odot$ and peculiar velocity limit of < 9 km s$^{-1}$, from a more recent general model[5,16] it would be inferred a SN mass loss < 1 $M_\odot$ (figure 5).

Unfortunately, from the first GAIA data release it was not possible to derive more precise parallax distances and proper motions of Cygnus X-1 and Cygnus OB3. New data release from GAIA may allow in few years' time a test of the hypothesis of BH formation by implosion and a more accurate constraint of a putative NK for the BH of Cygnus X-1.

**GRS 1915+105** is a low-mass X-ray binary containing a BH of $10.1 \pm 0.6$ $M_\odot$[23,24,25] and a donor star of spectral type K-M III of $0.5 \pm 0.3$ $M_\odot$ with a 34 day circular orbital period[23]. The companion overflows its Roche lobe and the system exhibits episodic superluminal radio jets[26].

Using a decade of astrometry of GRS 1915+105 with the NRAO Very Long Baseline Array and the published radial velocity, a modest peculiar velocity of $22 \pm 24$ km s$^{-1}$ has been reported[24], which is consistent with the earlier proposition[27] that the BH in GRS 1915+105 was formed without a strong NK, like the $14.8 \pm 1.0$ $M_\odot$ BH in Cygnus X-1. The new parallax distance implies that GRS 1915+105 is at about the same distance from the Sun as some compact

HII regions and water masers associated with high-mass star formation in the Sagittarius spiral arm[24]. Because those compact HII regions are located along the direction of powerful relativistic jets from GRS 1915+105 and show bow-shock structure, it was suggested[28] that massive star formation could have been induced by the interaction of jets and massive outflows from the BH-XRB with the interstellar medium at distances of ~50 parsecs from GRS 1915+105.

The modest peculiar speed of 22 ± 24 km s$^{-1}$ on the Galactic disk at the parallax distance[24], and a donor star in the giant branch suggest that GRS 1915+105 is an old system that has orbited the Galaxy many times, acquiring a peculiar velocity component on the galactic disk of 20-30 km s$^{-1}$, consistent with the velocity dispersions of ~20 km s$^{-1}$ of old stellar systems in the thin disk[29], due to galactic diffusion by random gravitational perturbations from encounters with spiral arms and giant molecular clouds.

### 2.2 Runaway black hole X-ray binaries

The physical mechanisms that may impart the runaway velocity of a BH are of topical interest to observationally constrain SN models and population-synthesis models of BH binary evolution. The runaway velocities of BH-XRBs can be caused by different physics mechanisms; a variety of mechanisms based on those originally proposed to explain the runaway massive stars, namely, baryonic mass loss in the sudden SN explosion of the primary star of a massive binary (Blaauw kick)[4], dynamical interactions in high density stellar environments[8,9], Galactic diffusion by random gravitational perturbations from encounters with spiral arms and giant molecular clouds for anomalous velocities up to 20-30 km s$^{-1}$, and BH NKs[6,7]. There are two types of BH NKs[30], those imparted intrinsically to the BH by asymmetric gravitational waves[7] and/or asymmetric neutrino emission[6,31] during core-collapse, and those imparted to the transient NS that turn into a BH by mass fallback.

Several efforts were recently undertaken to estimate BH NKs from the observations of low-mass BH-XRBs[32,33]. From the statistical analysis and model binary evolution of low mass BH-XRBs binaries with determined positions it has been proposed[34,32,5] that in order to achieve their distances from the Galactic disk, BHs may receive high NKs at birth[32]. This motivated the proposition based on theoretical calculations that by the gravitational pull from asymmetric mass ejecta, BHs can be accelerated to velocities comparable to those of NSs[35]. However, because of the unknown origin and several other uncertainties of the samples of sources, it has been argued that from only the existing observations of the spatial locations of low-mass X-ray binaries, it is not possible to confidently infer the existence of high BH NKs[36]. In the following are reviewed the observations of the three runaway low mass BH-XRBs for which the space velocities in three dimensions have been determined: GRO J1655-40, XTE 1118+480 and V404 Cyg.

**GRO J1655-40** is an X-ray binary with a BH of 5.3 ± 0.7 M$_\odot$ and a F6-F7 IV donor star with a runaway velocity of 112 ± 18 km s$^{-1}$ moving in a highly eccentric (e=0.34 ± 0.05) Galactic orbit[37]. The overabundance of oxygen and alpha-elements in the atmosphere of the donor star has been interpreted as evidence for SN ejecta captured by the donor star[38]. The runaway linear momentum of this X-ray binary is similar to those of solitary runaway neutron stars and millisecond pulsars with the most extreme runaway velocities.

It had been proposed[5] that the large runaway velocity of this BH-XRB may be accounted for by a successful SN in which a mass of ~4 $M_\odot$ was ejected, that a natal kick is not needed to explain its large space velocity, and that a fallback of 3 $M_\odot$ after the SN explosion would be required in order to explain the kinematics of the system[39]. In a more recent model[40] it is found that although a symmetric BH formation event cannot be formally excluded, the associated system parameters would be marginally consistent with the currently observed binary properties. It has been argued that BH formation mechanisms involving an asymmetric SN explosion with associated BH kick velocities of a few tens of km s$^{-1}$, may satisfy the constraints much more comfortably[40]. However, it should be mentioned that the generally assumed distance and SN origin of the overabundances of α-elements observed in this Galactic BH-XRB binary have been challenged[41,42].

**XTE J1118+480** is a high-galactic-latitude (l=157°.78, b=+62°.38) X-ray binary with a BH of 7.6 ± 0.7 $M_\odot$ and a 0.18 $M_\odot$ donor star of spectral type K7 V–M1 V, moving in a highly eccentric orbit around the Galactic center region, as some ancient stars and globular clusters in the halo of the Galaxy[43]. Because of the higher than solar abundances of nucleosynthetic products found in the donor star[44], it has been proposed that the BH was formed in the Galactic disk through an energetic SN. In this context, an asymmetric kick velocity of 183 ± 31 km s$^{-1}$ would have been required to change from a Galactic disk orbit to the currently high latitude observed orbit[45]. Furthermore, from the analysis of all the available information it was proposed[46] that the BH was formed with an explosive mass loss, and a possible NK at the time of core collapse. In Figure 3 is shown the galactic orbital path of this BH binary.

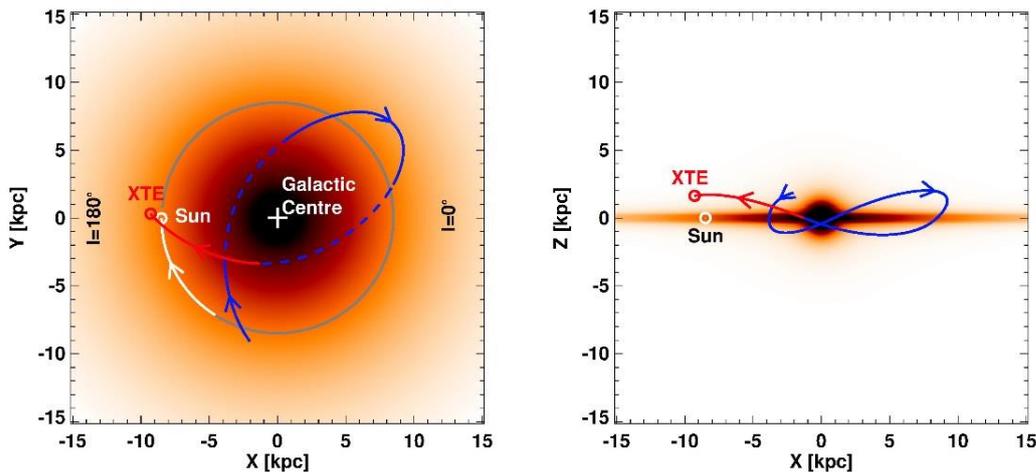

**Figure 3**. Schematic Galactic orbit of XTE J1118+480 (blue curve) during the last orbital period of the Sun around the Galactic Centre (240 Myr). The last section of the orbit since the source left the plane 37 ± 5 Myr ago at a galactocentric distance of 4 ± 0.5 kpc is shown in red. The trajectory of the Sun during the later time is indicated by the thick white arc. The source left the plane towards the northern Galactic hemisphere with a galactocentric velocity of 348 ± 18 km/s, which after subtraction of the velocity vector due to Galactic rotation, corresponds to a peculiar space velocity of 217 ± 18 km s$^{-1}$ relative to the Galactic disk frame, and a component perpendicular to the plane of 126 ± 18 km s$^{-1}$. The galactic orbit of XTE J1118+480 has an eccentricity of 0.54. At the present epoch XTE J1118+480 is at a distance from the Sun of only 1.9 ± 0.4 kpc flying through the Galactic local neighborhood with a velocity of 145 km.s$^{-1}$. **a**, View from above the Galactic plane; **b**, side view. (Adapted from ref 43)

**V404 Cyg (GS 2023+338)** is a low mass X-ray binary system composed of a BH of 9.0 ± 0.6 M$_\odot$ and a 0.75 M$_\odot$ donor of spectral type K0 IV. Using astrometric VLBI observations, it was measured[47] for this system a parallax that corresponds to a distance of 2.39 ± 0.14 kpc. Based on that fitted distance it was derived a peculiar velocity of 39.9 ± 5.5 km s$^{-1}$, with a component on the Galactic plane of 39.6 km s$^{-1}$, which is ~2 times larger than the expected velocity dispersion in the Galactic plane. Because of this relative mild anomalous motion it was proposed[48] that likely the peculiar velocity is due to a mass loss in a spherically symmetric supernova explosion, with any additional asymmetric kick being small.

In the atmosphere of the donor star of V404 Cyg, contrary to GRO J1655-40, the abundances of Al, Si, and Ti appear to be only slightly enhanced when comparing with average values in thin-disk solar-type stars, but the oxygen abundance is particularly enhanced[49]. Metal-rich spherical explosion models are able to reproduce the observed abundances relatively well and provide the energy required to explain the peculiar velocity of this system. In the context of this hypothesis[49], one would expect a milder kick –if any- in V404 Cyg, relative to that in GRO J1655-40.

**MAXI J1836−194** is an X-ray binary discovered in 2011[50]. Due to its radio and x-ray properties it is considered as a BH candidate. From VLBI astrometric observations of the compact radio counterpart, the proper motion for a source distance between 4 kpc and 10 kpc, and systemic radial velocity of 61 ± 15 km s$^{-1}$, imply a peculiar space velocity for this system of >70 km s$^{-1}$, from which it was concluded that the system required an asymmetric natal kick to explain its observed space velocity[51]. However, at present the uncertainties on the distance and mass of the compact object are still too large.

### 2.3 Discussion on the kinematics of black hole X-ray binaries

In Figure 4 are represented the barycenter's linear momenta of the X-ray binaries computed from the data in Table 1, as a function of the BH masses. A similar diagram with similar results was published before[12], but plotting in the abscissas the velocity instead of the linear momentum. Except XTE J1118+480, the other 4 sources are in the Galactic disk (b < 3.2°; z < 0.15 kpc) where likely were formed, and have peculiar motion components perpendicular to the Galactic plane lower that 6 ± 1 km s$^{-1}$. XTE J1118+480 is in the Galactic halo (b = 62.3°; z = 1.5 kpc) and could either have been propelled to its present position from the Galactic disk by an energetic BH natal event[44,45,46] or have been formed in a globular cluster from which it could have escaped[43] with a mild velocity of few tens km s$^{-1}$, preserving most of the globular cluster motion.

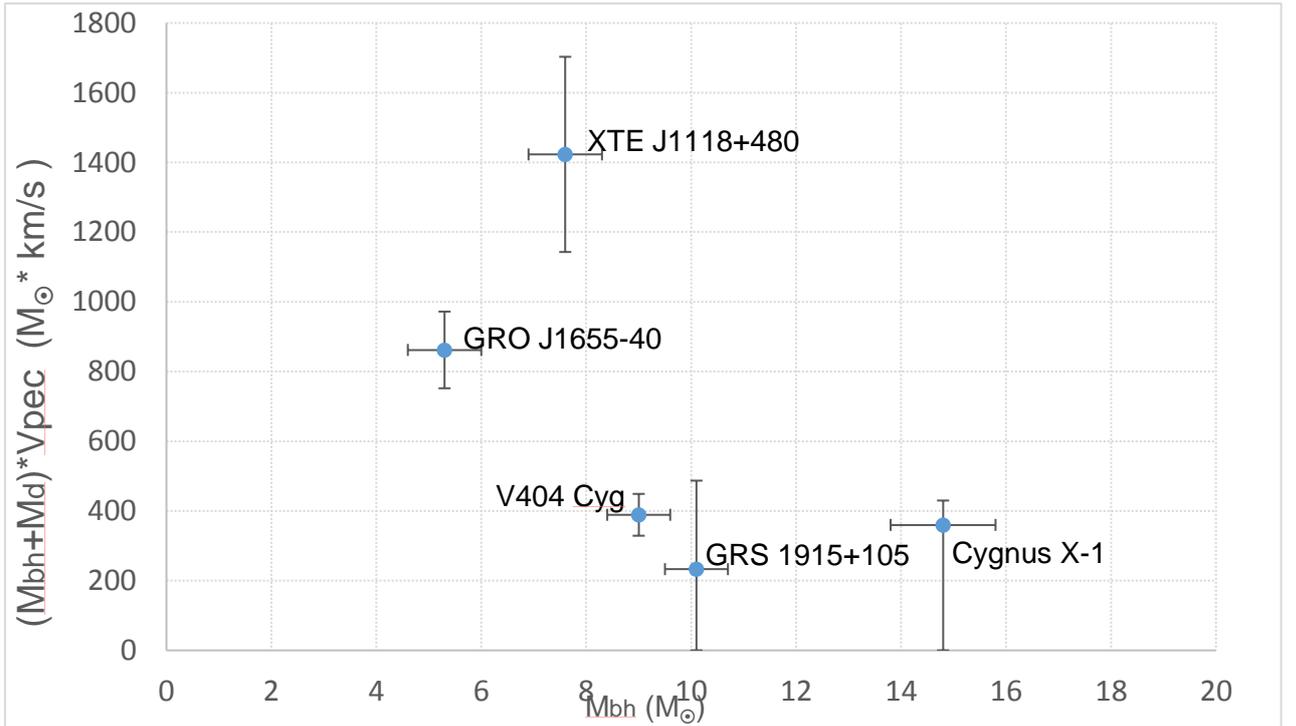

**Figure 4.** Linear momentum of the X-ray binaries as a function of the black hole masses. Despite the low numbers statistics of sources with velocities in the three dimensions of space, apparently there is an increase of the peculiar linear momenta of the X-ray black hole binaries with decreasing black hole mass, as expected from several models on black hole formation[1,22,13,53]. All sources are in the Galactic disk (b < 3.2°; z < 0.15 kpc) where likely were formed, except XTE J118+480 which is in the Galactic halo (b = 62.3°; z = 1.5 kpc) with an uncertain origin, either in the Galactic disk[44,45,46] or in a globular cluster[10]. If formed in a globular cluster, the large linear momentum of XTE J1118+480 relative to its last passage through the Galactic disk would have a large velocity component from the parent globular cluster and this BH-XRB should be removed from the figure.

Despite the low numbers statistics, Figure 4 shows a possible trend of increasing linear momenta of the BH-XRBs barycenter's as a function of decreasing BH mass, as expected from theoretical models[1,22,13,53], where higher mass stars are believed to end as BHs by direct collapse, whereas lower mass BHs are formed via SN fallback on a proto-neutron star. If the BHs in XRBs are formed by mass fallback on a NS, for linear momentum conservation, it is expected an anti-correlation between BH mass and NK; the larger fallback and BH mass, the lower the NK.

If the X-ray binaries in Figure 4 were formed from binaries in a field of relative low density, one would conclude that the three BHs with <10 M$_\odot$ were formed with significant baryonic ejections from SN explosions or NKs, whereas the BHs with >10 M$_\odot$ were formed by complete or almost complete collapse with no NKs. But if the runaway BH binaries were formed in dense stellar clusters, the anomalous velocities of the X-ray binaries barycenter's could have been caused by dynamical interactions in the stellar cluster, or by the explosion of a dynamically bound massive star, rather than by the explosion of the BH progenitor in the runaway binary. For instance, one possible scenario is that of a triple system composed by a BH low mass binary bound to a massive star that suddenly explodes. In this scenario the BHs of 5 M$_\odot$, 7 M$_\odot$, and 9 M$_\odot$ in the three runaway X-ray binaries discussed here, could have been formed by implosion with no natal triggers, before the later explosion of the third stellar component. In this context, the runaway velocities and SN nucleosynthetic products in the atmospheres of the donor stars in GRO J1655-40, XTE 1118+480 and V404 Cyg could be due to the explosion of a previously bound star, rather than to the explosion of the BH progenitor in the runaway binary.

The three runaway X-ray binaries have low mass donors, and their linear momenta are comparable to those of runaway massive stars ejected from multiple stellar systems by the Blaauw mechanism[4]. Therefore, without knowing the origin of a runaway X-ray binary, it is not possible to certainly constraint from its peculiar velocity alone, the strength of a putative NK to the compact object in the X-ray binary.

If the BHs in these sources were triggered at birth by Blaauw kicks or NKs, it would be intriguing that the components of the anomalous motions perpendicular to the Galactic disk of GRO J1655-40, V404 Cyg, GRS 1915+105 and Cygnus X-1 are respectively of only, 2.1±1, 4±1, 6±2, and 6±1 km s$^{-1}$, unless there is some unknown preference for kicks with directions contained in the Galactic disk. The most striking case is that of GRO J1655-40, for which was inferred a velocity component perpendicular to the Galactic plane, of only 2.1±1 km s$^{-1}$ whereas its anomalous velocity component on the Galactic plane is ~112±18 km s$^{-1}$, with a Galactic orbital eccentricity of 0.34±0.05, and reaching maximum distances from the plane of 50-150 pc.

It is expected that with GAIA at optical wavelengths and VLBI at radio wavelengths it will be possible to obtain a more precise parallax distance and proper motion for GRO J1655-40 and a larger sample of sources to constrain BH kicks.

**2.4 Black holes formed by implosion: observations versus theory**

The kinematics in three dimensions of Cygnus X-1 relative to the parent association of massive stars Cygnus OB3, and the kinematics of GRS 1915+105 relative to its Galactic environment, suggest –irrespective of their origin in isolated binaries or in dense stellar environments- that the BHs in Cygnus X-1 and in GRS 1915+105, were formed in situ by complete or almost complete collapse of massive stars, with no significant baryonic SN mass ejection and/or energetic NKs. These observational results are consistent with theoretical models[1,53,22], and in particular, with one of the most recent models[16], where the BHs of 14.8 $M_\odot$ in Cygnus X-1, and of 10.1 $M_\odot$ in GRS 1915+105, could have been formed by complete and/or almost complete collapse of helium cores.

In Figure 5 are shown the results from a recent model[16] of explosions for the core-collapse of solar metallicity stars of 9 to 120 $M_\odot$. The compactness of the stellar core is an important structural characteristic of a pre-supernova star to turn what is initially an implosion into an explosion[2,55,56]. A small compactness favors explosion whereas a large compactness favors implosion. Stars from 22 to 26 $M_\odot$ are hard to explode, between approximately 35 to 50 $M_\odot$ even harder to explode, and end as BHs[16].

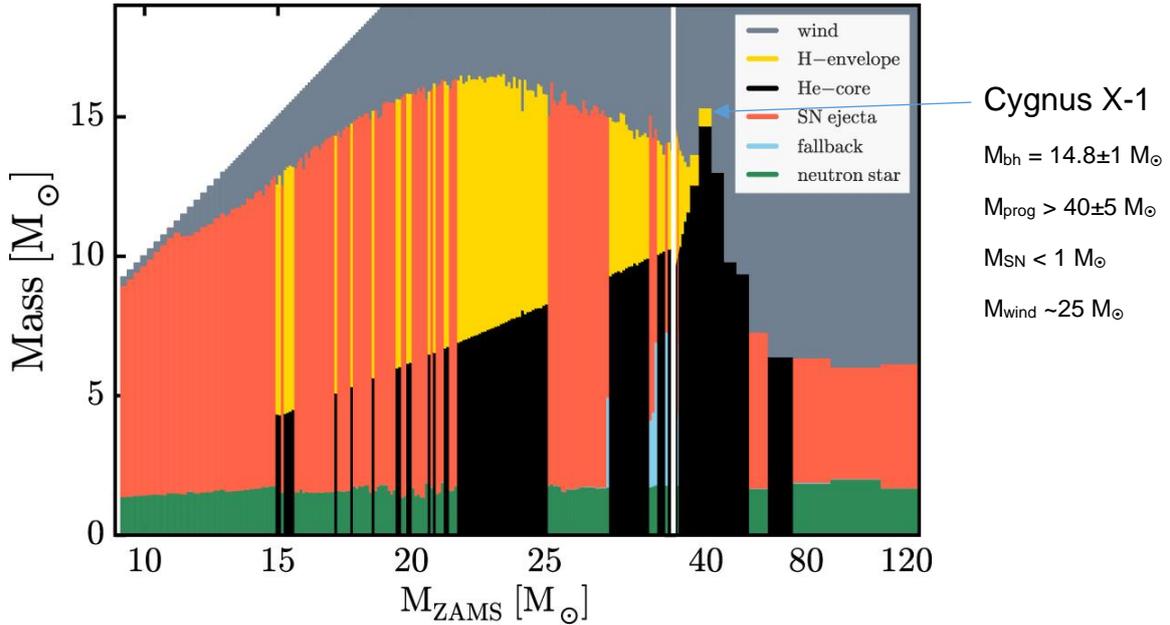

**Figure 5.** Core-collapse Supernovae from 9 to 120 Solar Masses. A novel result of this model relative to that shown in figure 1 is that for zero age main sequence stars (ZAMS) of solar metallicity with masses between ~15 and ~80 $M_\odot$, there are islands of exploitability (represented in red), in a sea of BH formation (represented in black). This model[16] of mass budget for ZAMS collapsing stars of solar metallicity suggests the formation of the BH in Cygnus X-1 by almost/complete collapse of the He core with an ejected H-envelope of < 1 $M_\odot$, as had been inferred from the X-ray binary kinematics in three dimensions[19]. For the successfully exploded cases, the compact remnant mass of neutron stars is shown in green. For a few models that experienced fallback, the fallback mass is shown in blue. Except for narrow ZAMS regions between ~30 and ~40 $M_\odot$, fallback is negligible. The helium cores and the hydrogen envelopes of the "failed" explosions are shown in black and yellow, respectively. The resulting BH mass from an implosion will most likely include the full pre-SN star (black plus yellow), or just the helium core (black). (Reproduced from ref. 16)

A novel aspect of this model is the use of a one-dimensional neutrino transport model for the explosion of stars along the lines of previous "neutrino powered models"[56,57]. In these models the iron core of a massive star collapses to a NS, depending on the critical neutrino heating efficiency required for exploding a neutron star progenitor with a given compactness structure[2]. The binding energy of the NS is radiated as neutrinos, a fraction of which deposit their energy in the matter above the NS causing it to expand and explode. Stellar collapses that fail to create a strong outward moving shock after 3-15 seconds form BHs, assuming that the core of helium and heavy elements collapses into the BH. The fate of the hydrogen envelope is less clear. In stars of solar metallicity and masses above ~30 $M_\odot$ the H envelope will already have been lost by winds. In the lighter stars, all of which are red supergiants, the envelope is very tenuously bound and any small core disturbance prior to explosion or envelope instability could lead to its ejection. Even if the envelope is still in place when the iron core collapses, the sudden loss of mass from the core as neutrinos can lead to the unbinding of the envelope[58].

In Figure 5 the masses of He cores are represented in black, the masses of NSs in green, the masses of the H-envelopes in yellow, those of SN ejecta in red, those lost by winds in grey and the fallback masses in blue, for the core-collapse of zero age main sequence stars of solar metallicity. The final masses of the BHs are those of the He cores represented in black plus an uncertain fraction of the outer H-envelope mass represented in yellow.

As inferred from the observations[19] presented in section 2.1, in this model on the evolution of single massive stars the BH of 14.8 ± 1.0 $M_\odot$ in Cygnus X-1 was formed by direct collapse of a star of ~40±5 $M_\odot$, with a baryonic mass ejection of <1 $M_\odot$ by a putative faint SN explosion, and in this model for single stars evolution, ~25 $M_\odot$ are lost by stellar winds.

The agreement of BH formation from single stars shown in figure 5 with the observations of the binary Cygnus X-1 may be surprising, unless possible early interactions, mass transfer and common envelope phase in the evolution of Cygnus X-1 have been irrelevant for the core collapse of the primary star. Alternatively, a chance coincidence due to uncertainties in the observations and/or theoretical model cannot be excluded.

The progenitor of the BH of 10.1 ± 0.6 $M_\odot$ in GRS 1915+105 is more uncertain. It could have been formed from a star of zero age main sequence between ~35 and ~60 $M_\odot$ by complete or almost complete collapse of a He core of ~9-10 $M_\odot$, but the mass loss by winds is uncertain.

### 2.5 Stellar black holes in globular clusters

It is expected that in a typical globular cluster (GC) with present day masses of $10^5$–$10^6$ $M_\odot$, hundreds of stellar-mass BHs should be born during the first ~10 Myr after formation. The specific frequency of X-ray binaries in GCs is ~100 times larger than in the field and provides strong evidence that in GCs mass-transferring binaries are dynamically formed with high efficiency.

If BHs are present in GCs, a subset should be detectable as accreting binaries[59]. However, it is known that the majority of the most luminous X-ray binaries in GCs are accreting NSs since they are detected as sources of Type I X-ray bursts, which are thermonuclear explosions on the hard surface of accreting NSs.

In the last decade BH candidates may have been identified in galactic and extragalactic GCs. It was assumed that these sources are candidate BH binaries because they have X-ray luminosities above the Eddington luminosities of NSs, and they vary significantly on short timescales, making it implausible that the luminosity could come from a superposition of several NS X-ray binaries. The most luminous ULXs so far identified are associated with a GC in the massive Virgo elliptical galaxy NGC 4472[60], which has a peak X-ray luminosity $L_X \sim 4 \times 10^{39}$ erg s$^{-1}$. Optical spectroscopy of the associated GC exhibits broad (1500 km s$^{-1}$) [OIII] emission but no Balmer lines.

The few BHs possibly observed in extragalactic GCs would be those with either the most extreme accretion rates or very massive. They likely represent only the very tip of the iceberg in terms of BH X-ray binaries in GCs. Many BHs with lower accretion rates are almost certain to exist among X-ray sources in GCs, but they are greatly outnumbered by NS binaries and are difficult to distinguish from NSs using X-ray data alone[59].

Because ULXs may also be powered by NSs[61], a new strategy for identifying quiescent BH X-ray binaries in Milky Way GCs makes use of both radio and X-ray data. Stellar-mass BHs accreting at low rates have compact jets which emit radio continuum via partially self-absorbed synchrotron emission. BHs are much more luminous in the radio than NSs with comparable X-ray luminosity; in fact $L_R/L_X$ is ~2 orders of magnitude higher for BHs than NSs.

Before the recent upgrade to the VLA, the radio emission from a quiescent BH like V404 Cyg would not have been detectable at high significance at typical GC distances. The upgraded VLA can now readily detect the expected flux densities (tens of µJy) in reasonable exposure times. Using observations at radio frequencies two candidate stellar mass BHs in the core of the Galactic GC cluster M 22 have been found[62]. These sources have flat radio spectra and 6 GHz flux densities of 55–60 µJy. As these sources are not detected in shallow archival Chandra imaging, they cannot yet be placed directly on the $L_X$–$L_R$ relation; nevertheless, their overall properties are consistent with those expected from accreting BH binaries. Another BH candidate in a second Galactic GC, M62 (NGC 6266; D = 6.8 kpc) named M62-VLA1 was discovered[59]. Unlike the former cases for the M22 sources, M62-VLA1 has clear X-ray and optical counterparts, and so it is the most compelling candidate BH X-ray binary in a Milky Way GC.

It was believed that BHs formed by implosion in GCs fall to the center, where accreting BH X-ray binaries as XTE J1118+480[43] and BH–BH binaries like GW150914 may then be formed through three-body interactions[10] that lead to recoil velocities much larger than the escape velocities from typical GCs of few tens of km s$^{-1}$. However, it has been shown recently[63] that core collapse driven by BHs (through the Spitzer "mass segregation instability") is easily reverted through three-body processes, and involves only a small number of the most massive BHs, while lower mass BHs remain well-mixed with ordinary stars far from the central cusp, suggesting that stellar BHs could still be present today in large numbers in many GCs.

### 2.6. Intermediate mass black holes in globular clusters

The formation process of supermassive BHs is still uncertain and one of the possible clues to understand their origin may reside in the evolutionary connection between stellar mass and supermassive BHs through the formation of intermediate mass BHs (IMBHs) of $10^2$ to $10^4$ $M_\odot$. Although it is believed that such objects should be formed in dense stellar systems such as GCs, the observational evidences for their existence had been elusive. For instance, based on the lack of electromagnetic counterparts in X-rays[64] and radio waves[65], in the GC 47 Tucanae, which is at distance of ~4 kpc, upper limits of 470 and 2,060 $M_\odot$ had been placed on the mass of a putative IMBH.

Recently, probing the dynamics of the GC 47 Tucanae with pulsars it has been inferred[66] the existence of a gas-starved IMBH with a mass of 2,200 (+1,500/-800) $M_\odot$. The authors conclude that this BH is electromagnetically undetectable due to the absence of gas in the core within the radius of influence of the IMBH, and that IMBHs as the one in 47 Tuc may constitute a subpopulation of progenitor seeds that formed supermassive BHs in galaxy centers.

## 3. BH formation as a function of metallicity and redshift

Theoretical models on the evolution of single massive stars[67,13,15,68] predict mass and metallicity dependence of stellar progenitors for BH formation (e.g. Figure 1). For stellar clusters of a given mass it has been proposed that the numbers and orbital period distributions of HMXBs should also depend on metallicity[69]. More recent models[70] predict with respect to previous models, significantly larger values of the carbon–oxygen core mass of massive stars, which would imply substantially larger BH masses at low metallicity (≤ 2 x 10$^{-3}$), than previous population synthesis codes. According to this model[70] the maximum BH masses for a given progenitor mass would be ~25, 60 and 130 $M_\odot$ for metallicity progenitors of $Z/Z_\odot$ = 2 x 10$^{-2}$, 2 x 10$^{-3}$ and 2 x 10$^{-4}$.

On the other hand, hydrodynamic simulations on the formation of the first generations of stars in the universe, show that a substantial fraction of stars in primordial galaxies are formed in small, multiple, top-heavy Initial Mass Function groups, with a high incidence of binaries with typical masses of several tens of solar masses[71,72,73]. In fact, more than 70% of stars of spectral type O in the Galaxy are in binaries, and massive binaries in the Milky Way have flat mass ratios[74] (e.g. half of the companions of a 40 $M_\odot$ star have > 20 $M_\odot$).

In the context of these models, where BH formation depends on metallicity and can take place by direct and failed SN core-collapse, it was proposed that BH-HMXBs, namely, the fossils of the first generations of binary massive stars, should have been prolifically produced at cosmic dawn[75,76,77]. At present this hypothesis cannot be contrasted observationally at very high redshifts.

However, the metallicity dependence of BH-HMXBs has been observationally confirmed for galaxies in the local universe[78,79,80]. Using a large set of data on the sizes and X-ray luminosities (mostly coming from accreting BHs) of HMXBs in nearby galaxies with known metallicities and star formation rates, it is found that HMXBs are typically ten times more numerous per unit star formation in low-metallicity galaxies (<20% solar) than in solar-metallicity galaxies[80].

The expected redshift dependence of BH formation on metallicity has been observationally confirmed[81], and more recently by observations in the Chandra Deep Field South survey[82]. The X-ray luminosity due to HMXBs in normal galaxies show out to z=2.5 a redshift evolution given by

$$L_{\text{2-10 keV}} (\text{HMXB})/\text{SFR} \propto (1 + z)$$

which is mostly due to the declining metallicity of the progenitors of HMXBs with increasing redshift[82]. Extrapolations of the results of this survey suggest that at $z > 6$, HMXBs in normal galaxies would produce an X-ray emissivity that exceeds that of AGN[82].

Therefore, in the context of these theoretical models and recent observational results it is expected that at high redshifts BBHs should have been formed more readily than in the Local Universe. However, due to the multiple uncertainties on the different BBH formation channels and common envelope transfer in massive stellar binaries[53], a quantitative estimation of the frequency of BBH formation as a function of redshifts beyond z=2.5 at this time is rather uncertain and beyond the scope of this observational review.

## 4. Progenitors of BHs: massive stars that collapse in the dark?

Models of the evolution of massive stars had predicted that stars between ~25 and 140 $M_\odot$ and up to solar metallicity end as BHs by direct or failed supernova collapse[6,13,22] (figure 1). More recent models predict that zero age main sequence (ZAMS) stars of solar metallicity with masses between 15 and 80 $M_\odot$ may end as BHs, but in some intervals of ZAMS masses, stars may explode as SNe and end as NSs[16] (figure 5). In this section are reviewed observations at optical and infrared wavelengths that seem to be consistent with the predictions of these models.

### 4.1. Absence of very massive progenitors of Type IIP supernovae

Type IIP supernovae and their progenitors make up to 40 % of all SN explosions in the local Universe. It is believed that type IIP supernovae are the explosions of red supergiant stars (RSGs) with masses between 10 and 40 $M_\odot$ that have retained most of their hydrogen envelopes throughout their evolution, which results in a ~100d hydrogen recombination plateau in the light curve. Theory[83] predicts that RSG progenitors of masses between 8 and 25 $M_\odot$ should explode.

One approach to identify the progenitors of Type IIP SNe in nearby galaxies is by means of high resolution archival images from space and ground-based telescopes. Among 45 Type IIP SNe with either detected progenitors or upper limits, it was found[84] a remarkable deficit of stars above an apparent limit of log $L/L_\odot$ ~5.1 dex which was translated into a mass limit of 16.5-18.5 $M_\odot$ much lower of the theoretical limit of 25 $M_\odot$. This apparent inconsistency between theory and observations has been called the "RSG problem".

Figure 6 illustrates the absence of luminous RSGs progenitors with estimated masses > 16.5-18.5 $M_\odot$, from which it has been suggested[84] that the bulk of stars above that upper mass limit may end as BHs with no visible SN. Furthermore, with the lack of detected high mass progenitors in type IIb SNe, it has been proposed that this "missing high mass problem" has become relevant to all type II SNe[84]. Based on recent observations[85] that are reviewed in section 4.5, it has been proposed that those observational low luminosity and mass limits relative to those theoretically expected, are a consequence of the underestimation of absorption due to the limited information provided by the archival images used in this approach.

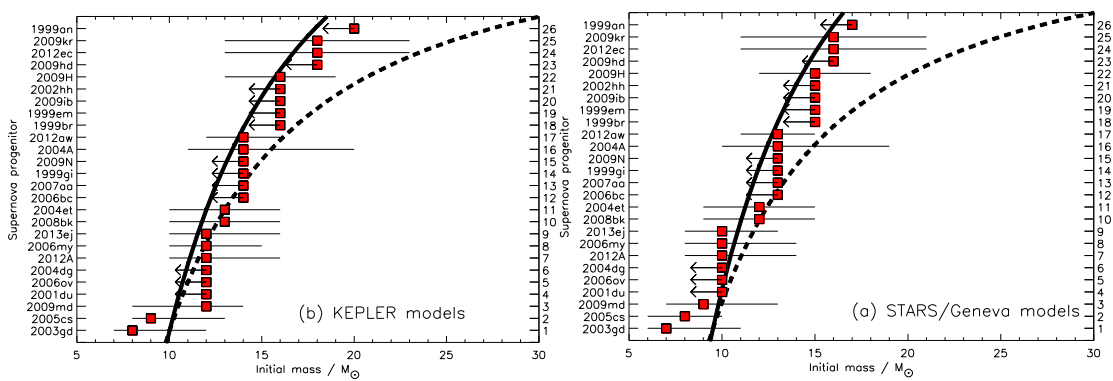

**Figure 6.** Mass of stellar progenitors of core collapse SNe in the context of the KEPLER[56] and STARS[45]/ Geneva[88] models of stellar evolution. The detections are marked with error bars, the limits with arrows, and the lines extend from the minimum to the maximum masses from cumulative Salpeter IMFs. Allowing the mass function to vary up to 30 $M_\odot$, the mass distribution would need to be truncated at masses of ~18.5 $M_\odot$ and ~16.5 $M_\odot$ respectively. Reproduced from ref. 84.

## 4.2 Massive stars that disappear without optically bright SNe

Repeated observations of luminous and massive stars in nearby galaxies reveal massive stars that disappear quietly, without optically bright supernovae[86]. Following this strategy, from a systematic analysis of archival Hubble Space Telescope (HST) images of 15 galaxies[89] and a survey of 27 galaxies with the Large Binocular Telescope[90], were found respectively, one yellow supergiant candidate of 25-30 $M_\odot$ that underwent an optically dark core-collapse (figures 7a,b), and one red supergiant candidate[90] with an estimated mass of ~25 $M_\odot$, likely associated with failed supernovae (figures 8a,b)

Fig. 7a shows the existence of some extended unresolved flux at the location of the "NGC3021-CANDIDATE-1", which is located in an arm of the spiral galaxy. As massive stars tends to be found in star-forming complexes, this extended emission may be the host cluster of NGC3021-CANDIDATE-1. If this is the case, then NGC3021-CANDIDATE-1 clearly dominated the flux of the cluster and was its most massive member, which would be consistent with it being the next star to collapse[89].

**Figure 7a**                                                                                          **Figure 7b**

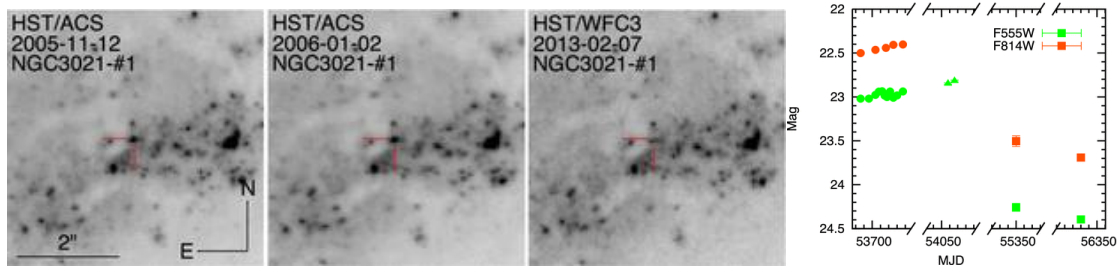

**Figure 7:**

a) Selected HST F814W (λ 802.4 nm) image cuts centered on the position of a candidate in the galaxy NGC3021, the location of which is indicated with red tick marks. Between 2005 and 2013 were taken in total 23 images. The first HST image was taken in 2005, some 3.5 years later, NGC3021-CANDIDATE-1 appeared to be ~1.5 mag fainter in F555W (λ 5407 nm) and ~1 mag fainter in F814W (λ 802.4 nm). A final set of observations in early 2013 show the source to be similarly faint. The region of the candidate shows several partially resolved extended sources that may be a host cluster of NGC3021-CANDIDATE-1. If this is the case, then NGC3021-CANDIDATE-1 dominated the flux of the cluster and was its most massive inhabitant, which would be consistent with it being the next star to collapse. It is estimated that this candidate is a 25–30 $M_\odot$ yellow supergiant that underwent an optically dark core-collapse.

b) Light curve of the candidate NGC 3021 at F555W (λ 5407 nm) and F814W (λ 802.4 nm). Whether the source detected at late times is related with the candidate in NGC 3021 is not clear. Alternatively, it could be related to some extended background flux or to the host cluster of the candidate. (Reproduced from ref. 89).

Recently, it was confirmed[91] by HST imaging the optical disappearance of a failed SN candidate, previously identified with the Binocular Telescope[90]. This ~25 $M_\odot$ RSG experienced a weak $10^6$ $L_\odot$ optical outburst in 2009 and until 2015 remained at optical wavelengths at least ~5 magnitudes fainter than the progenitor. In Figure 8 are shown the mutiwavelength observations of the failed supernova candidate now called N6946-BH1[91].

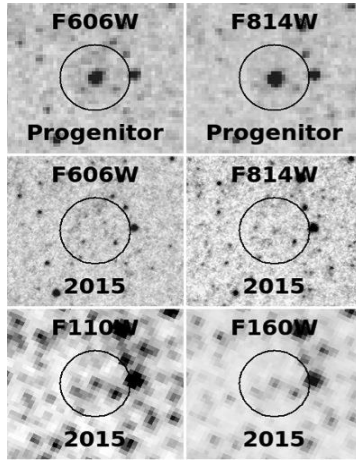
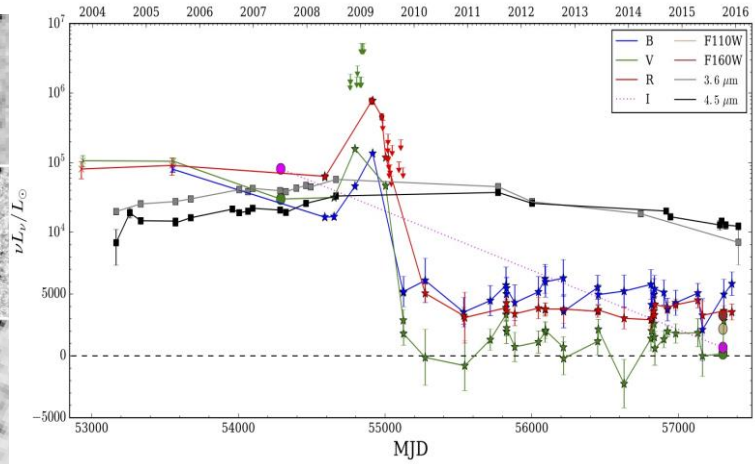

**Figure 8a:** HST images of the region surrounding N6946-BH1. The top and middle rows show the WFPC2 F606W (λ 588.7 nm) and F814W (λ 802.4 nm) images. The bottom row shows the WFC3/IR F110W (λ 1153.4 nm) and F160W (λ 1536.9 nm) images. The circles have a radius of 1". The progenitor has dramatically faded in the optical but there is still faint near-IR emission. Reproduced from ref. 91

**Figure 8b:** N6946-BH1 light curves from HST (large circles), Spitzer Space telescope (SST, squares), Large Binocular Telescope (LBT, stars), CFHT (x's), Palomar Transient Factory (red pentagons and upper limits), and amateur astronomer Ron Arbour (green upper limits). The vertical axis switches from a linear scale below $10^4$ $L_\odot$ to a logarithmic scale above $10^4$ $L_\odot$. A luminosity of zero is given by the dashed black line for comparison. The uncertainties for the differential LBT and SST photometry do not include the significant uncertainty in their "zero points" created by crowding. The LBT fluxes at late times could easily be zero. For this purpose, the high resolution HST constraints that any remaining optical flux is < $10^3$ $L_\odot$ are more relevant. Reproduced from ref. 91.

Figures 7b and 8b show in both candidates remaining faint near infrared emission after more than 6 years since a faint transient display. In addition, figure 11b shows in the region of N6946-BH1 that the mid-infrared emission measured with Spitzer at 3.6μm and 4.5μm slowly decreased to the lowest levels since the first measurements, which has been interpreted[91] as radiation from dust formed in the material ejected at few hundreds of km s$^{-1}$ during outburst.

The estimated masses of 20-25 $M_\odot$ for the two stellar progenitors are in the mass range of the missing RSG SN progenitors[84] described in section 4.1, and in the high core "compactness" identified in theoretical models as most likely to give rise to failed SNe and direct BH formation[16]. The light curves in figures 7b and 8b are consistent with the low-energy, long-duration, red events theoretically predicted in BH formation by failed SN[58]. In this model[58] of a RSG collapse, a very weak SN with total kinetic energy of ~$10^{47}$ erg is produced, a large fraction of the hydrogen envelope is ejected with speeds of a few 100 km s$^{-1}$, and luminosities ~$10^{39}$ erg s$^{-1}$ are maintained for a year from hydrogen recombination.

Alternatively, it has been proposed[91,92] that this near-infrared late-time emission may be due to fallback accretion onto a newly formed BH obscured by dust formed in the weakly-ejected envelope. If this late-time luminosity is powered by fallback accretion, X-rays could be detected with Chandra, as long as the neutral hydrogen column depth is not too large[91]. Then, according to the authors[91] a detection of X-rays could lead to the first observational evidence of BH formation in almost real time. If confirmed, N6946-BH1 would be the first failed SN and first black hole birth ever discovered, and the problem of the missing high-mass SN progenitors would be solved[91].

It should be pointed out that near-infrared emission has been detected from accreting BHs in the Milky Way, but identified as radiation from synchrotron jets. Near-infrared emission from accreting BHs has been observed from jets associated to mayor soft X-ray outbursts[93] and at low accretion rates with time variabilities having periods of minutes[94], and as faster infrared flickering with time variations up to seconds[95].

In this context, an alternative way to confirm BH formation at the position of disappearing massive stars would be by recurrent observations at centimeter radio wavelengths, which are less affected than X-rays by the expected large HI column densities produced in failed SNe of core-collapse massive stars. If a recently formed BH accretes mass from matter falling back from the failed SN, or from stellar companions in a multiple bound stellar system (e.g. a binary companion), feedback in the form of time variable relativistic jets are likely to be produced. Milky Way BHs accreting at high rates produce giant recurrent jets of up to tens of Jy's at radio wavelengths (e.g. Cygnus X-3[96] & GRS 1915+105[97]). At the distance of ~6 Mpc of NGC 6946 analogous giant outbursts would be detected with the upgraded VLA with flux densities in the range of 1 to tens µJy's.

Given the detection of 6 successful SNe in the sample of originally 27 monitored galaxies with the Large Binocular Telescope and one likely failed SN, the implied fraction of core-collapses that result in failed SNe is f=0.14 +0.33/-0.10 at 90% confidence, and if the current candidate is ultimately rejected, there is a 90% confidence upper limit on failed SN fraction of f<0.35[92].

### 4.3. Absence in color–magnitude plots of SN progenitors with >20 M$_\odot$

An indirect strategy was applied[98] using resolved stellar photometry from archival HST imaging to generate color–magnitude diagrams of stars within 50 pc of the location of 17 historic core-collapse SNe that took place in galaxies within a distance of 8 Mpc. Fitting the color–magnitude distributions with stellar evolution models to determine the best-fit age distribution of the young population, the authors conclude that so far there is not a single high-precision measurement of a SN progenitor with >20 M$_\odot$. But the authors point out that the large uncertainties for the highest-mass progenitors also allow the possibility of no upper-mass cutoff, hinting that there could be a ceiling to SN production or a mass range that under produces SNe.

### 4.4 Absence of spectral signatures for >20 M$_\odot$ progenitors of Type IIP SNe

SN 2012aw is a nearby Type IIP SN with a potentially luminous and massive progenitor that was directly identified in pre-explosion images. The inferred luminosity of the progenitor was disputed in several separate analyses, but is potentially higher than most previous RSGs and a candidate to be the first progenitor star of a Type IIP SN with a ZAMS >20 $M_\odot$.

Besides photometry at different wavelengths, the nucleosynthetic products in the nebular spectra of SNe can also provide constraints on the mass of the exploding star. Combining the individual evolution in the three oxygen lines excited by thermal collisions ([O I] $\lambda$5577, [O I] $\lambda$6300, and [OI] $\lambda$6364) the mass of the oxygen present in the carbon burning ashes can be constrained.

Using optical and near-infrared nebular spectroscopy between 250 and 451 days after the explosion[99] it was realized that the observed evolution of the cooling lines of oxygen in the nebula are difficult to reconcile with the expected nucleosynthesis products from progenitors of Type IIP SNe with > 20 $M_\odot$.

Reviewing[99] the literature of published nebular spectra of Type IIP SNe were found no observations where the [O I] $\lambda$6300, $\lambda$6364 lines are significantly stronger (relative to the optical spectrum as a whole) than in SN 2012aw, from which the authors[99] conclude that no Type IIP SN has yet been shown to eject nucleosynthesis products from stars with masses above 20 $M_\odot$.

### 4.5. Discussion on the high mass limit for progenitors of Type IIP SNe

It has been argued that the optical and infrared observational approaches described above, and the conclusions inferred from them on the upper mass-limit of core-collapse SNe that lead to the "RSG problem" may have several biases, as those possibly due to the influence of circumstellar dust, luminosity-mass analysis, magnitude variations of RSGs, sample selection, and limited numbers statistics.

Recent measurements of circumstellar dust in 19 RSGs of the coeval cluster NGC 2100 showed that the mass lost rate significantly increases through the lifetime of RSGs before exploding as SNe, with mass loss rates a factor of 40 higher close to explosion than in early stages[85]. In this study it was also found evidence for an increase of circumstellar extinction through the RSG lifetime, implying that in the more evolved stars the progenitor's initial mass could be underestimated by up to 9 $M_\odot$, in which case the observational upper mass limits of 16-20 $M_\odot$ for the progenitors of Type IIP SNe that had been inferred from archival images, when corrected by absorption, would be consistent with the theoretical mass limit of 25 $M_\odot$, and the "RSG problem" solved.

Under the assumption of mass lost during RSGs evolution, current models[100] predict that RSGs will evolve back to bluer regions of the magnitude-color diagram, the heaviest RSGs moving to the yellow supergiant phase where may end their lives as BHs with little or no explosion at all. Interestingly, one of the two most likely candidates of BH formation by failed SNe identified so far[89] with an estimated initial mass between 25 and 30 $M_\odot$ (section 5.2) is a yellow supergiant.

## 5. The formation of binary stellar black holes

The first sources of gravitational waves detected by LIGO were mergers of stellar BHs[101] and the question on how these BBHs may be formed is of topical interest. In previous sections have been presented observational results that are consistent with the idea that stellar black holes may be formed by the implosion of massive stars in the dark, without luminous natal SNe, with no ejection of a significant amount of baryonic matter, and with no energetic kicks.

### 5.1. BBHs formed by isolated evolution of massive stellar binaries

BH high mass X-ray binaries (BH-HMXBs) are an evolutionary stage of isolated massive stellar binaries, before BBH formation (figure 9).

Massive Stellar Binary      BH-HMXB      Binary black hole

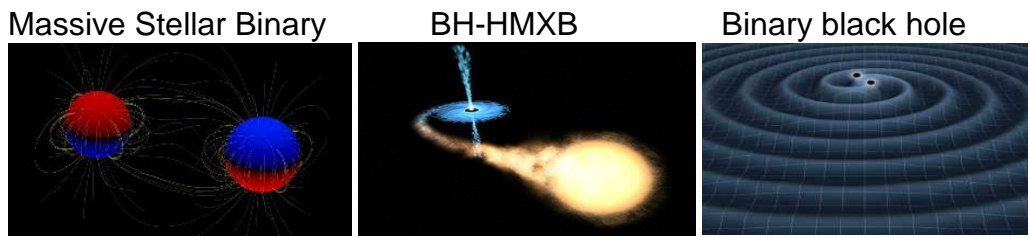

**Figure 9.** Binary black holes may be formed from relatively isolated massive stellar binaries through an intermediate phase of a black hole high mass X-ray binary (BH-HMXB). For metallicities $Z < 0.1\ Z_\odot$ stars of $M > 20\ M_\odot$ may collapse directly and form BBHs. Massive stars of solar metallicity $Z = Z_\odot$ lose mass by stellar winds and may also collapse directly as black holes for certain values of the Zero Age Main Sequence mass of the progenitor ($\sim > 40\ M_\odot$ for the black hole in Cygnus X-1, see section 3.1). It has been shown that it is unlikely that the Galactic BH-HMXB Cygnus X-1 will become a BBH[30]; perhaps Cygnus X-3, but it is not guaranteed[102].

The BHs that merged producing the source of gravitational waves GW150914[103] had masses of ~30 $M_\odot$, much larger than the 5 $M_\odot$ to 15 $M_\odot$ stellar BHs found so far in the Milky Way, and it has been proposed[67] that those BHs were formed by direct collapse. In Figure 10a is shown an example of a field binary evolution leading to a BH-BH merger similar to GW150914[67]. This model invokes mass transfer from the secondary to the first BH during common envelope, which still is a poorly understood evolutionary phase of BH-HMXRBs that leads to large uncertainties.

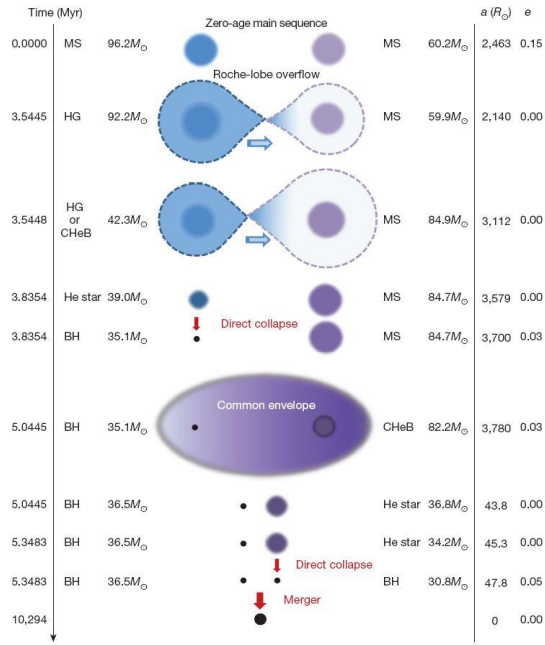

**Figure 10a.** Example of the evolution of a massive stellar binary leading to a BH–BH merger similar to GW150914[67]. A massive binary star (96 $M_\odot$ (blue) + 60 $M_\odot$) is formed in the distant past (2 billion years after Big Bang; $z \approx 3.2$; top row), and after 5 million years of evolution forms a BH–BH system (37 $M_\odot$ + 31 $M_\odot$; second-last row). For the ensuing 10.3 billion years, this BH–BH system is subject to loss of angular momentum, with the orbital separation steadily decreasing, until the black holes coalesce at redshift $z = 0.09$. This example binary formed in a low-metallicity environment ($Z = 0.03\ Z_\odot$) where the BHs were formed by direct collapse. MS, main-sequence star; HG, Hertzsprung-gap star; CHeB, core-helium-burning star; BH, black hole; $a$, orbital semi-major axis; $e$, eccentricity. Reproduced from reference 67.

## 5.2. BBHs formed from tight binaries with fully mixed chemistry

Another scenario is that of a massive over contact binary (MOB) that remain fully mixed as a result of their tidally induced spins. This chemically homogeneous evolutionary channel for BBH formation in tight massive binaries[104,105] is insensitive to kicks smaller than the binary's orbital velocity. In this model BBHs originate from binaries in or near contact at the onset of hydrogen burning, which allows mixing of the burning products in the center throughout the stellar envelope, a process originally proposed for rotating single stars[106]. At low metallicities MOBs will produce BBHs that merge within Hubble time with mass ratios larger than 0.55, as in GW150914. The schematic representation in Figure 10b shows that this model avoids the physics uncertainties in common envelope ejection events, and the still unconstrained BH Kicks. This channel has a preference for high total and chirp masses as in GW150914, but BBHs with the relative low masses of ~10 $M_\odot$ of the BHs as in GW151226[107] are difficult to reproduce.

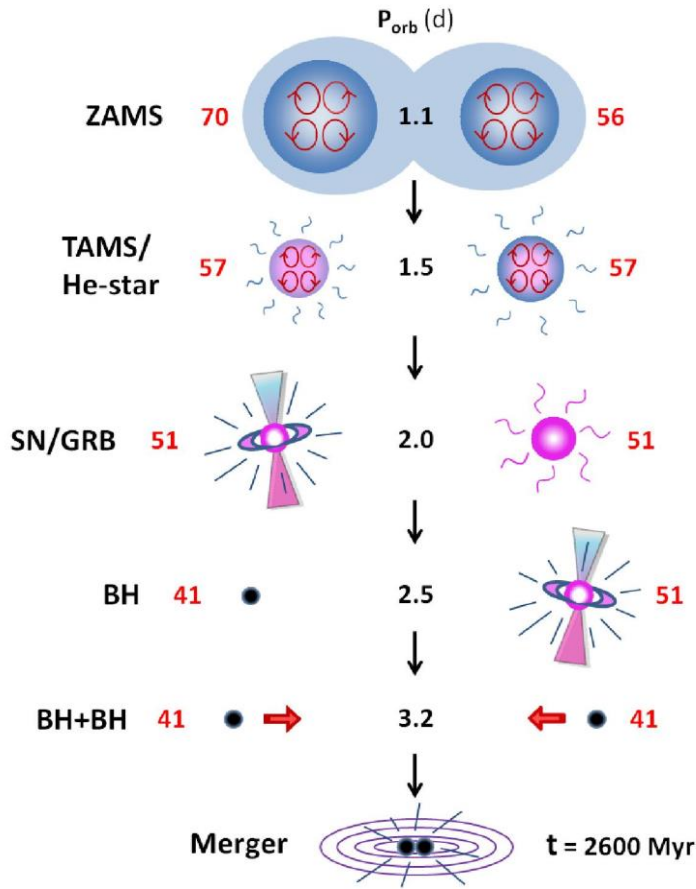

**Figure 10b.** Illustration of the binary stellar evolution leading to a BH+BH merger with a high chirp mass in the context of the "massive over-contact binary evolution[105]" also called "chemical homogeneous evolutionary channel for BH mergers[104]". This model avoids the physics uncertainties in common envelope ejection events, and the still unconstrained BH Kicks.

The initial metallicity is $Z_\odot/50$, the masses of the stars in solar masses are indicated with red numbers, and the orbital periods in days are given as black numbers. A phase of contact near the ZAMS causes mass exchange. Acronyms used in the figure: ZAMS: zero-age main sequence; TAMS: termination of hydrogen burning; He-star: helium star; SN: supernova; GRB: gamma-ray burst; BH: black hole. Reproduced from reference 105.

### 5.3. BBHs formed by dynamical interaction in dense star clusters

In Figure 10c are shown alternative paths for the formation of the BBH progenitor of GW150914 by dynamical interactions in GCs[10]. These paths for BBH formation assume that members of the BBH are formed with no energetic SNe or NKs that would disrupt the binary system or eject the BH components out from the cluster before BBH formation. The escape velocity from a typical GC is a few tens of km s$^{-1}$ and BHs with kick velocities of hundreds of km s$^{-1}$ as observed in some NSs would be ejected from typical GCs, unless BBHs are preferentially formed by dynamical interactions in nuclear clusters of $10^7$ M$_\odot$ or more with sizes of only a few parsecs[108]. However, due to uncertainties in the conversion of luminosity to mass, the actual existence and frequency of such super-massive clusters is a matter of debate. For instance, from HST observations of nuclear clusters in the starburst galaxies M82 and NGC 5253 it has been estimated[109,110] that the most massive nuclear clusters in these nearby galaxies have typical sizes of a few parsecs but masses between 7 x $10^4$ M$_\odot$ and up to 1.3 x $10^6$ M$_\odot$. Anyway, it is known that large numbers of NSs have been retained in GCs and must have been born with low or no overall kicks.

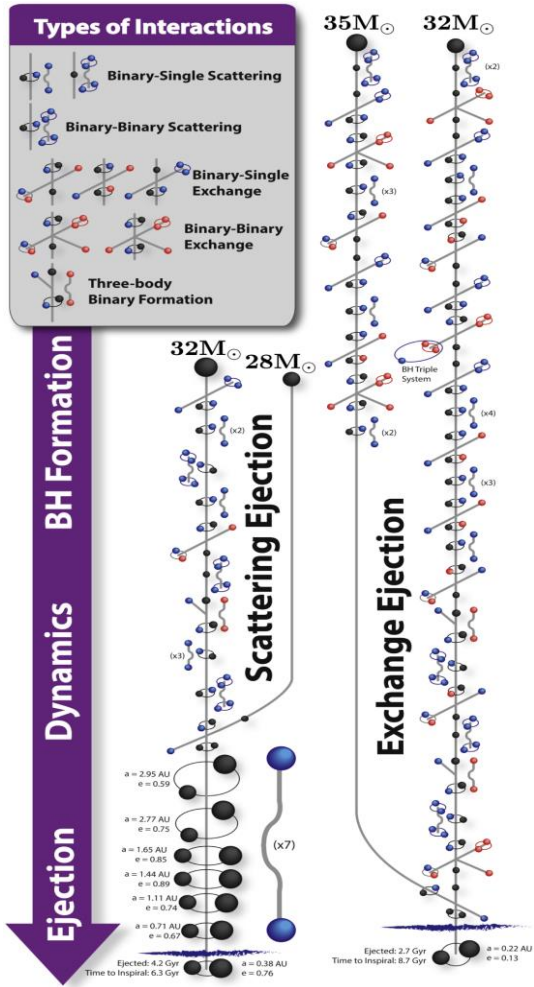

**Figure 10c.** Dynamical Formation of the GW150914 Binary Black Hole in a Globular Cluster. This interaction diagram shows the formation history for two GW150914 progenitors in a single GC model. From top to bottom, the history of each individual BH that will eventually comprise a GW150914-like binary is illustrated, including all binary interactions. The legend shows the various types of gravitational encounters included in these GC models (with the exception of two-body relaxation). In each interaction, the black sphere represents the GW150914 progenitor BH, while the blue and red spheres represent other BHs (and stars) in the cluster core. Reproduced from Ref. 10.

**5.4. GW150914 & GW151226 versus theoretical BBH formation paths**

Figure 11 shows the posterior probability densities for the masses of the sources components of the three GW events identified so far[101]. The probable high masses (~30 $M_\odot$) of the components in GW150914 and their mass ratios larger than 0.55 are consistent with BBHs formed from tight binaries with fully mixed chemistry (model 5.2). However, the relative low masses (~10 $M_\odot$) of the BH components in GW151226 suggest that the BBH of this event was either formed from an isolated binary (model 5.1), or by dynamical interaction in a high stellar density cluster (model 5.3), with the BHs formed without kicks that would unbind the isolated binary (model 5.1), or eject the BHs from the parent stellar cluster before BBH formation (model 5.3). The likely low kicks at birth and ~10 $M_\odot$ of the BHs in GW151226 are qualitatively consistent with the observational evidence for in situ formation of BHs of similar masses in Galactic XRBs (section 2.1).

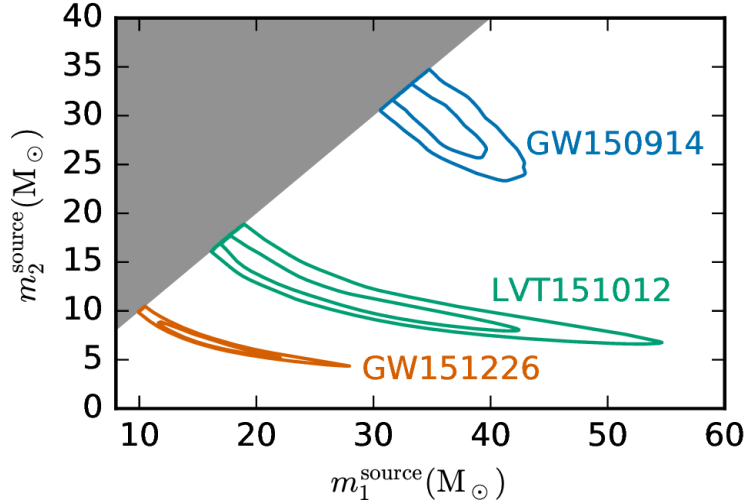

**Figure 11**. Posterior probability densities for the masses of the sources components of the three events GW150914, LVT151012 and GW151226. For the two dimensional distributions, the contours show 50% and 90% credible regions. Reproduced from reference 101.

## 6. Summary and conclusions

- From the kinematics of Galactic BH X-ray binaries in three dimensions of space it is found that stars of solar metallicity and >40 M$_\odot$ may collapse directly to form BHs by implosion, without energetic SN explosions and Natal Kicks (NKs). In fact, from the kinematics of Cygnus X-1 it is inferred that the BH of ~15 M$_\odot$ was formed in situ with no NK, by the implosion of a progenitor of ~40 M$_\odot$, that probably went through a Wolf Rayet phase with a total mass loss of ~25 M$_\odot$. The kinematics of GRS 1915+105 suggests that the BH of ~10 M$_\odot$ in this X-ray binary was also formed by implosion with no trigger from a NK. These observations are consistent with core-collapse models based on neutrino-powered explosions of massive stars (Sections 2.1 and 2.4).

- The linear momentum of BH X-ray binaries determined from velocities in three dimensions increases with decreasing BH mass, as expected from core-collapse models. However, this possible trend is inferred from low numbers statistics (only five sources) and its interpretation in terms of kick velocities imparted to the compact object in the runaway binary is uncertain without knowledge of the binary's origin. X-ray binaries may be formed in different environments and their runaway velocities be caused by a diversity of physical mechanisms (sections 2.2 and 2.3).

- XTE J1118+480 is in the Galactic Halo and if interpreted as a field binary formed in the disk would have received a natal kick >80 km s$^{-1}$, but if formed by dynamical interaction in a globular cluster the BH could have been formed with no energetic trigger. The BH X-ray binaries GRO J1655-40, V404 Cyg, GRS 1915+105 and Cygnus X-1 are in the Galactic disk and have relatively low motions in directions perpendicular to the disk. They are respectively 2.1±1, 4±1, 6±2, and 6±1 km s$^{−1}$. If the BHs in those X-ray binaries would have been born with energetic kicks, it would be intriguing that their motions in directions perpendicular to the Galactic disk are so small, unless there is a preference of BH natal kicks in directions along the Galactic disk…The most striking case is that of the runaway BH X-ray binary GRO J1655-40, which has a velocity component on the Galactic plane of ~112±18 km s$^{-1}$ whereas its velocity component perpendicular to the Galactic plane is 2.1±1 km s$^{-1}$ (section 2.3).

- It is expected that parallax distances and proper motions of BH-XRBs determined from VLBI observations at radio wavelengths, and with GAIA at optical wavelengths, will allow to determine the velocities in three dimensions accurately enough and for larger samples of BHXRBs, to track their paths to the sites of birth, and better constrain models on stellar BH formation.

- BH-XRB candidates have been identified in globular clusters, and their confirmation would provide evidence for BH formation by either direct collapse and/or sufficiently low natal kicks, since they would not have been ejected from the clusters. It is expected that low mass BHs accreting at low rates could still be present in large numbers far from the globular cluster's central cusps. A way to identify those quiescent BH X-ray binaries in Milky Way globular clusters is by the observation of the characteristic radio continuum self-absorbed synchrotron emission of stellar BHs at low mass accretion rates.

- The recent possible identification of a dormant intermediate-mass BH candidate in the globular cluster 47 Tucanae inferred from the dynamical state of the cluster, opens a new way to identify these elusive objects, which could have been seeds for the formation of supermassive BHs. (section 2.6).

- The theoretically expected metallicity and redshift dependence of the formation of BH-XRBs has now been confirmed by observations. From a large set of data of high mass X-ray binaries (mostly containing accreting BHs) in nearby galaxies, it is found that they are typically ten times more numerous per unit star formation in low-metallicity galaxies (<20% solar) than in solar-metallicity galaxies. The expected redshift dependence of high mass X-ray binaries on metallicity has also been confirmed by observations in the Chandra Deep Field South survey. The X-ray luminosity normalized by the star formation rate due to BH-HMXBs in normal galaxies show out to redshift z=2.5 an evolution that is proportional to redshift, which is due to the declining metallicity of the progenitors of BH-HMXBs with increasing redshift (section 3). It is expected that the future X-ray satellite Athena will allow to extend the study of BH-XRBs to higher redshifts.

- A large fraction of stars above some mass limit subject to debate (between 17 and 25 M$_\odot$) implode in the dark without luminous SNe, most likely ending as BHs. This is inferred from searchers for SN progenitors in optical and infrared archived images, from massive stars that quietly disappear in the dark, from the largest stellar masses in the young stellar populations that host historic supernova remnants, and from the absence of nucleosynthetic products of very massive stars in the nebular spectra of core-collapse supernovae. The detection of accreting BHs at the position of failed SNe by follow up observations in X-rays and radio wavelengths, offers the possibility to find the first observational evidence of BH formation in real time. The enhanced capabilities of the future Athena X-ray satellite and the upgraded VLA and future SKA radio interferometers will play an important role in this area of research (section 4).

- Three main evolutionary channels have been proposed for Binary Black hole (BBH) formation: (1) BBHs formed by isolated evolution of massive stellar binaries, (2) BBHs formed from tight binaries with fully mixed chemistry, and (3) BBHs formed by dynamical interaction in dense stellar clusters. In model (1) the BH members of BBHs are formed by direct collapse and with no BH natal kicks that would unbind the stellar binary. In model (3) it is tacitly assumed that the members of BBHs are also formed with no BH natal triggers that would eject the BHs from the stellar cluster before BBH formation. Model (2) avoids the physics uncertainties in mass transfer, common envelope mass ejection events, and the still unconstrained BH kicks. However, model (2) assumes massive tight binary progenitors of BBHs and has preference for BBHs with large masses as in GW150914, but the relative low BH masses in GW151226 are difficult to reproduce. In principle the formation of the BBH in GW151226 can be accounted by models (1) and (3). (Section 5).

- BBHs formed from relatively isolated massive stellar binaries (channel 1) or contact massive binaries (channel 2) by either direct collapse and sufficiently low natal kicks, will remain in situ and ultimately merge in galactic disks. BBHs formed by dynamical interactions in the cusps of globular clusters, may be ejected from their birth place and ultimately merge in galactic haloes, like the sources of short gamma-ray bursts.

- The detection by the LIGO-Virgo collaboration of GWs from the fusion of stellar BBHs has open new horizons for BH astrophysics. Virgo will soon come into operations, and together with LIGO and other future GW detectors will make possible to narrow down the origin of the GWs, allowing the detection of possible electromagnetic radiation coming from the same GW source or its immediate environment, and the identification of its astronomical host. It is expected that in a more distant future the GW space mission LISA, besides detecting GWs from the fusion of supermassive BHs, will be able to anticipate the time at which the final fusion of stellar BHs will take place, allowing extraordinary progresses in the frontiers of physics and astrophysics.

- Most astrophysical insights on the formation of stellar BHs by implosion of massive stars had been based on what we don't see: e.g. the absence of runaway motions in BH-XRBs, the absence of progenitors of core-collapse SNe above some mass limit, the absence of luminous SNe associated to massive stars that suddenly disappear, the absence in the nebular spectra of core-collapse SNe of the expected nucleosynthetic elements from very massive stars …
Thanks to the faith on scientific research and technological creativity that made possible the detection of the first sources of gravitational waves, now we have seen direct signals from BHs. Perhaps Saint Augustine was right when in another context he stated: "Faith is to believe what you do not see, the reward of this faith is to see what you believe."

**Acknowledgements**: Thierry Foglizzo, Luis F. Rodríguez, Françoise Combes and James Miller-Jones made insightful comments to the first draft of this review. An anonymous referee made important remarks to the submitted version that helped to provide a more appropriate balance to the presentation of this review. Selma de Mink, Tassos Fragos, Garik Israelian, Smadar Naoz, Jerome Orosz, Jorge Melnick and Jochen Greiner provided useful information. Irapuan Rodrigues and Stéphane Schanne kindly helped with software to produce figures. This work was supported by the National Programs PNHE and PNCG of the French Institut National des Sciences de l'Univers.

**Table 1. Parameters of black hole X-ray binaries and peculiar velocities**

| BH-XRB | $M_{BH}$ ($M_\odot$) | $M_{donor}$ ($M_\odot$) | Sp Type | $V_{pec}$ (km s$^{-1}$) | e (galactic) | e (orbital) | P (days) |
|---|---|---|---|---|---|---|---|
| GRO J1655-40 | 5.3±0.7[1,16,19] | 2.4±0.7[1,19] | F6-7IV[1,19] | 112±18[2] | 0.34±0.05[2] | 0.0[19] | 2.62[1] |
| XTE J1118+480 | 7.6±0.7[3] | 0.5±0.3[3] | K7-M1 V[5] | 183±31[4,18] | 0.54±0.05[18] | 0.0[14] | 0.17[6] |
| V 404 Cyg | 9.0±0.6[7] | 0.75±0.25[8] | K0 IV[7] | 39.9±5.5[9] | 0.16±0.02[15] | 0.0[8] | 6.47[8] |
| Cygnus X-1 | 14.8±1.0[10] | 19.2±1.9[10] | O9.7Iab[10] | < 9±2[11] | | 0.018±003[10] | 5.6[10] |
| GRS 1915+105 | 10.1±0.6[20] | 0.5±0.3[20] | K-M III[12] | 22±24[13] | 0.28±0.05[17] | 0.0[12] | 34[12] |

**Table 1.** BH-XRB = Black hole X-ray binary with velocities determined in three dimensions, $M_{BH}$ = Mass of black hole in solar units, $M_{donor}$ = Mass of donor star, Sp Type = Spectral Type of donor star, $V_{pec}$ = Peculiar velocity of the BH-XRB barycenter relative to its birth site or environment, e (galactic) = Galactic orbital eccentricity, e (orbital) = binary orbital eccentricity, P = binary orbital period.